\documentclass[prl,twocolumn,showpacsb]{revtex4}
\usepackage{graphicx}
\usepackage{dcolumn}
\usepackage{bm}

\input{tcilatex}

\begin{document}

\title{Swapping entangled Kondo resonances in parallel-coupled double
quantum dots}
\author{Guang-Ming Zhang$^{1,2}$, Rong L\"{u}$^{1}$, Zhi-Rong Liu$^{2}$, and
Lu Yu$^{3}$}
\affiliation{$^1$Center for Advanced Study, Tsinghua University, Beijing 100084, China;\\
$^{2}$Department of Physics, Tsinghua University, Beijing 100084, China;\\
$^{3}$Interdisciplinary Center of Theoretical Studies and Institute of
Theoretical Physics, CAS, Beijing 100080, China.}
\date{\today}

\begin{abstract}
Strong electron and spin correlations are studied in parallel-coupled double
quantum dots with interdot spin superexchange $J$. In the Kondo regime with 
\textit{degenerate} dot energy levels, a coherent transport occurs at zero
temperature, where two entangled (bonding and antibonding) resonances are
formed near the Fermi energy. When increasing $J$ or the dot-lead
parallel-coupling \textit{asymmetry} ratio $\Gamma _{2}/\Gamma _{1}$, a swap
between two entangled resonances occurs and the line shapes of the linear
conductance are interchanged. The zero-bias differential conductance shows a
peak at the critical values. Such a peculiar effect with the virtue of
many-body coherence may be useful in future quantum computing.
\end{abstract}

\pacs{73.21.La, 73.23.-b, 73.63.Kv}
\maketitle

A large number of proposals have been made to materialize quantum bits
(qubits) and quantum computing. Among these proposals, coupled quantum dot
(QD) systems are particularly attractive.\cite{loss-1998} The spin degree of
freedom of the localized electrons on the dots is considered as a qubit due
to the comparatively long coherence time. A key challenge is the
construction of coupled double QD (DQD) to perform a swap operation, \textit{%
i.e.} exchanging the electron spin states on the two dots. When the square
root of a swap operation is combined with other isolated qubit rotations, a
quantum controlled-NOT gate can be built and any quantum algorithms can be
implemented.\cite{divincenzo-1995} In the coupled DQD, two local electrons
form a singlet state. It has been proposed that the swap operation can be
realized by tuning the time-dependent interdot spin superexchange (ISS) $%
J(t) $ from positive to negative, flopping the singlet and triplet states.%
\cite{loss-1998,loss-1999}

In this Letter, we propose a simple and reliable mechanism to perform such a
swap process in a parallel-coupled DQD at low temperatures. It has been
well-established that under the Coulomb blockade with odd number electrons
on a single QD, a quantum coherent many-body (Kondo) resonance is formed
near the Fermi energy in the dot density of states (DOS) \cite{goldhaber}.
The Kondo effect for even number electrons on a single multilevel dot and
possible phase transitions between singlet and triplet states have been
considered for both ``vertical''\cite{vertical} and ``lateral'' \cite%
{hs-2002,hz-2003,pgh-2003,ka-2002} configurations. For the coupled DQD with
degenerate energy levels, the Kondo behavior and the ISS interplay and
strongly compete, as seen in the bulk two-impurity Kondo problem,\cite{jones}
and a question arises whether there exists a spin entangled state composed
of the coherent Kondo resonances.\cite{holleitner} Previous experimental and
theoretical studies have mainly focused on the serial-coupled DQD \cite%
{oosterkamp,georges,langreth,aono-eto,izumida-sakai,busser,dong,lopez},
except for Ref. \onlinecite{ka-2002,lopez}. However, it has been recently
realized that the serial geometry is \textit{unsuitable} for studying this
competition experimentally and a direct evidence for observing spin
entanglement between the dot local electrons could be sought in the parallel
coupled configuration.\cite{chen-chang-melloch} It has thus motivated us to
investigate whether and how this competition manifests itself in the
coherent transport through DQD in the parallel configuration.

\begin{figure}[tbp]
\begin{center}
\includegraphics[width=1.5in]{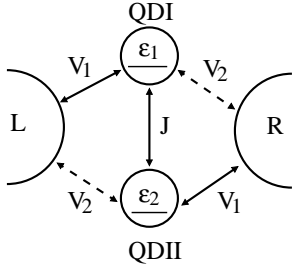}
\end{center}
\caption{Configuration of the coupled double quantum dots with asymmetric
couplings to the left and right electrodes. Two degenerate dot levels ($%
\protect\epsilon _{1}=\protect\epsilon _{2}$) are coupled by an AF spin
superexchange with strong on-site Coulomb repulsion.}
\label{configuration}
\end{figure}

For a degenerate DQD with ISS, a special dot-lead coupling configuration
(see Fig.~\ref{configuration}) is considered. By increasing the dot-lead
coupling asymmetry ratio $\Gamma _{2}/\Gamma _{1}$ ($\Gamma _{1,2}=\pi \rho
_{f}V_{1,2}^{2}$), where $\rho _{f}$ is the DOS at the Fermi level and $%
V_{1,2}$ is the dot-lead hopping integrals, the dot local electrons couple
to the right and left electrodes can be transformed from serial ($\Gamma
_{2}=0$) to symmetric parallel-coupled configuration ($\Gamma _{2}=\Gamma
_{1}$). We generalize the slave-boson mean field (MF) theory \cite{coleman}
to take into account both electron and spin correlations \textit{%
simultaneously}. Using an entanglement order parameter (EOP) $\Delta _{f}$
and the asymmetry ratio $\Gamma _{2}/\Gamma _{1}$ to describe the
entanglement between the local electron spins, we find that two entangled
bonding and antibonding Kondo resonances are formed very close to the Fermi
energy. When increasing the ISS or the dot-lead coupling asymmetry ratio,
the EOP can change sign at a critical value. As a result, the bonding and
antibonding resonances swap, or equivalently, the singlet and triplet levels
interchange and the zero-bias differential conductance displays a peak at
the critical value. Importantly such a swap effect occurs \textit{only} when
the dot-lead parallel couplings are \textit{asymmetric}, \textit{i.e}. $%
0<\Gamma _{2}/\Gamma _{1}<1$ and the gate voltage controlling the interdot
electron hopping is fine tuned.

We describe the coupled DQD system as two Anderson magnetic impurities with 
\textit{degenerate} levels and infinite on-site Coulomb repulsion, and an
antiferromagnetic(AF) spin superexchange between two local electron spins is
generated by the second order perturbation in the interdot electron
tunneling. The model Hamiltonian is given by 
\begin{eqnarray}
H &=&\sum_{\mathbf{k},\sigma ;\alpha =L,R}\epsilon _{\mathbf{k,\alpha }}C_{%
\mathbf{k},\sigma ,\alpha }^{\dagger }C_{\mathbf{k},\sigma ,\alpha
}+\sum_{\sigma ;i=1,2}\epsilon _{d}f_{i,\sigma }^{\dagger }f_{i,\sigma } 
\nonumber \\
&&+\frac{1}{\sqrt{N}}\sum_{\mathbf{k},\sigma }\left[ \left( V_{1}f_{1,\sigma
}^{\dagger }b_{1}+V_{2}f_{2,\sigma }^{\dagger }b_{2}\right) C_{\mathbf{k}%
,\sigma ,L}\right.  \nonumber \\
&&+\left. \left( V_{2}f_{1,\sigma }^{\dagger }b_{1}+V_{1}f_{2,\sigma
}^{\dagger }b_{2}\right) C_{\mathbf{k},\sigma ,R}+h.c.\right]  \nonumber \\
&&+J\left( 2\mathbf{S}_{1}\cdot \mathbf{S}_{2}+\frac{1}{2}\right) ,
\end{eqnarray}%
where $N$ is the total number of electrons, the slave-boson representations $%
d_{i,\sigma }=b_{i}^{\dagger }f_{i,\sigma }$ have been used to describe the
local electrons on each QD ($b_{i}$ and $f_{i,\sigma }$ denote the
respective hole and electron occupied states), and the local constraints $%
b_{i}^{\dagger }b_{i}+\sum_{\sigma }f_{i,\sigma }^{\dagger }f_{i,\sigma }=1$
have to be imposed \cite{coleman}. $\mathbf{S}_{1}$\textbf{\ }and\textbf{\ }$%
\mathbf{S}_{2}$ correspond to the spin density operators of the dot local
electrons, characterized by $S_{i}^{\alpha }=\sum_{\sigma ,\sigma ^{\prime
}}f_{i,\sigma }^{\dagger }\tau _{\sigma ,\sigma ^{\prime }}^{\alpha
}f_{i,\sigma ^{\prime }}$ with $\tau ^{\alpha }$ ($\alpha =x,y,z$) the Pauli
matrices. Here, we are only concerned with the equilibrium ($\epsilon _{%
\mathbf{k},L}=\epsilon _{\mathbf{k},R}$) properties. A similar coupling
configuration of the DQD model has been studied in the noninteracting case.%
\cite{guevara} 
Since the ISS interaction can be written in an SU(2) singlet form \cite%
{affleck} 
\[
J\sum_{\sigma ,\sigma ^{\prime }}f_{1,\sigma }^{\dagger }f_{1,\sigma
^{\prime }}f_{2,\sigma ^{\prime }}^{\dagger }f_{2,\sigma }=-J\sum_{\sigma
,\sigma ^{\prime }}:f_{1,\sigma }^{\dagger }f_{2,\sigma }f_{2,\sigma
^{\prime }}^{\dagger }f_{1,\sigma ^{\prime }}: 
\]%
in the MF approximation, an EOP $\Delta _{f}=\sum_{\sigma }\langle
f_{1,\sigma }^{\dagger }f_{2,\sigma }\rangle $ can be introduced to describe
the spin singlet between the local electrons on two separate dots. As in the
usual treatment of the Anderson impurity model at $T=0$, the bosonic
operators are replaced by their expectation values and the local constraints
by the Lagrangian multipliers $\lambda _{i}$. It has been established that
the slave-boson MF treatment captures the basic Kondo physics for the
single-impurity Anderson model at low temperatures, and such a theory
becomes exact for large local spin degeneracy.\cite{coleman} In the presence
of degenerate dot energy levels, we expect to have $\lambda _{1}=\lambda
_{2}=\lambda $ and $b_{1}=b_{2}=b_{0}$, namely, the degeneracy of the dot
energy levels can \textit{only }be\textit{\ }lifted by the ISS interaction.
Thus, an effective model Hamiltonian is obtained%
\begin{eqnarray}
H_{e} &=&\sum_{\mathbf{k},\sigma ;\alpha }\epsilon _{\mathbf{k}}C_{\mathbf{k}%
,\sigma ,\alpha }^{\dagger }C_{\mathbf{k},\sigma ,\alpha }+(\widetilde{%
\epsilon }_{d}-J\Delta _{f})\sum_{\sigma }\alpha _{\sigma }^{\dagger }\alpha
_{\sigma }  \nonumber \\
&&+(\widetilde{\epsilon }_{d}+J\Delta _{f})\sum_{\sigma }\beta _{\sigma
}^{\dagger }\beta _{\sigma }+2\lambda \left( b_{0}^{2}-1\right) +J\Delta
_{f}^{2}  \nonumber \\
&&+\frac{(\widetilde{V}_{1}+\widetilde{V}_{2})}{\sqrt{2N}}\sum_{\mathbf{k}%
,\sigma }\left[ \alpha _{\sigma }^{\dagger }\left( C_{\mathbf{k},\sigma
,L}+C_{\mathbf{k},\sigma ,R}\right) +h.c.\right]  \nonumber \\
&&+\frac{(\widetilde{V}_{1}-\widetilde{V}_{2})}{\sqrt{2N}}\sum_{\mathbf{k}%
,\sigma }\left[ \beta _{\sigma }^{\dagger }\left( C_{\mathbf{k},\sigma
,L}-C_{\mathbf{k},\sigma ,R}\right) +h.c.\right] ,  \nonumber
\end{eqnarray}%
where $\widetilde{\epsilon }_{d}=\epsilon _{d}+\lambda $ and $\widetilde{V}%
_{i}=b_{0}V_{i}$ are two renormalized parameters, and two canonical modes of
bonding and antibonding have been introduced by $\alpha _{\sigma }=\frac{1}{%
\sqrt{2}}\left( f_{\mathbf{1,}\sigma }+f_{\mathbf{2,}\sigma }\right) $ and $%
\beta _{\sigma }=\frac{1}{\sqrt{2}}\left( f_{\mathbf{1,}\sigma }-f_{\mathbf{%
2,}\sigma }\right) $. The difference between the bonding and antibonding
energies are just given by $2J\Delta _{f}\equiv J_{eff}$. Actually, $J_{eff}$
also corresponds to the effective energy splitting of the spin singlet and
triplet states formed by the renormalized local electrons on the dots. Thus,
a singlet-triplet transition can occur when the EOP $\Delta _{f}$ changes
sign, instead of reversing the sign of $J$.

When a Nambu spinor $\Phi _{\sigma }^{\dagger }=\left( f_{1,\sigma
}^{\dagger },f_{2,\sigma }^{\dagger }\right) $ is defined, the Fourier
transform of the retarded Green's function $-\langle T_{\tau }\Phi _{\sigma
}(\tau )\Phi _{\sigma }^{\dagger }(\tau ^{\prime })\rangle $ can be derived
as 
\[
\mathbf{G}_{f}(\omega )=\frac{\left[ \omega -\widetilde{\epsilon }_{d}+i(%
\widetilde{\Gamma }_{1}+\widetilde{\Gamma }_{2})\right] -\left[ J\Delta
_{f}+2i\sqrt{\widetilde{\Gamma }_{1}\widetilde{\Gamma }_{2}}\right] \sigma
_{x}}{\left[ \omega -\widetilde{\epsilon }_{\alpha }+i\widetilde{\Gamma }%
_{\alpha }\right] \left[ \omega -\widetilde{\epsilon }_{\beta }+i\widetilde{%
\Gamma }_{\beta }\right] }, 
\]%
where $\sigma _{x}$ is the Pauli matrix, and the DOS on each QD is given by $%
A_{f}(\omega )=\frac{1}{2\pi }\left[ \frac{\widetilde{\Gamma }_{\alpha }}{%
(\omega -\widetilde{\epsilon }_{\alpha })^{2}+\widetilde{\Gamma }_{\alpha
}^{2}}+\frac{\widetilde{\Gamma }_{\beta }}{(\omega -\widetilde{\epsilon }%
_{\beta })^{2}+\widetilde{\Gamma }_{\beta }^{2}}\right] $, corresponding to
a superposition of the bonding and antibonding resonances lying at energies $%
\widetilde{\epsilon }_{\alpha ,\beta }=(\widetilde{\epsilon }_{d}\mp J\Delta
_{f})$ with renormalized hybridization widths $\widetilde{\Gamma }_{\alpha
,\beta }=b_{0}^{2}\pi \rho _{f}\left( V_{1}\pm V_{2}\right) ^{2}$. Actually,
these are two entangled resonances with many-particle coherence, and we will
refer to them as entangled bonding and antibonding Kondo resonances. At the
symmetric parallel-coupling $\Gamma _{2}=\Gamma _{1}$, only the bonding
combination of the conduction electrons, i.e. $\left( C_{\mathbf{k},\sigma
,L}+C_{\mathbf{k},\sigma ,R}\right) $ couples to the localized electrons,
and the antibonding combination $\left( C_{\mathbf{k},\sigma ,L}-C_{\mathbf{k%
},\sigma ,R}\right) $ are completely dropped out. \cite{ka-2002}

To determine the MF order parameters $b_{0}^{2}$ and $\Delta _{f}$, the
corresponding self-consistent equations $\sum_{\sigma }\langle f_{i,\sigma
}^{\dagger }f_{i,\sigma }\rangle =1-b_{0}^{2}$ and $\sum_{\sigma }\langle
f_{1,\sigma }^{\dagger }f_{2,\sigma }\rangle =\Delta _{f}$, are rewritten as

\begin{eqnarray}
1-b_{0}^{2} &=&\frac{1}{\pi }\left[ \tan ^{-1}\left( \frac{\widetilde{\Gamma 
}_{\alpha }}{\widetilde{\epsilon }_{\alpha }}\right) +\tan ^{-1}\left( \frac{%
\widetilde{\Gamma }_{\beta }}{\widetilde{\epsilon }_{\beta }}\right) \right]
, \\
\Delta _{f} &=&\frac{1}{\pi }\left[ \tan ^{-1}\left( \frac{\widetilde{\Gamma 
}_{\alpha }}{\widetilde{\epsilon }_{\alpha }}\right) -\tan ^{-1}\left( \frac{%
\widetilde{\Gamma }_{\beta }}{\widetilde{\epsilon }_{\beta }}\right) \right]
,
\end{eqnarray}%
where the first equation represents the total quasiparticle occupations on
bonding and antibonding energy levels, while $\Delta _{f}$ corresponds to
the difference between these two occupations as follows from the Friedel sum
rule. Moreover, a third self-consistent equation from $\langle \partial
_{\tau }b_{i}\rangle =\langle \left[ b_{i},H\right] \rangle =0$ is needed,
yielding

\begin{equation}
\widetilde{\epsilon }_{d}-\epsilon _{d}=\frac{1}{\pi b_{0}^{2}}\left[ 
\widetilde{\Gamma }_{\alpha }\ln \frac{D}{T_{\alpha }}+\widetilde{\Gamma }%
_{\beta }\ln \frac{D}{T_{\beta }}\right] ,
\end{equation}%
where two characteristic energy scales are formally defined as $T_{\alpha
(\beta )}=\sqrt{\left( \widetilde{\epsilon }_{d}\mp J\Delta _{f}\right)
^{2}+(b_{0}^{2}\pi \rho _{f})^{2}\left( V_{1}\pm V_{2}\right) ^{4}}$ with $D$
as the bandwidth of the conduction electrons in leads. We have to stress
that these two energy scales can not be directly related to the Kondo
temperatures defined in the multilevel QD system.\cite{hs-2002,ka-2002} To
find the saddle point solution numerically, we choose $\Gamma _{1}=1$ as the
energy unit and $D=100$.

\begin{figure}[tbp]
\begin{center}
\includegraphics[width=3.4in]{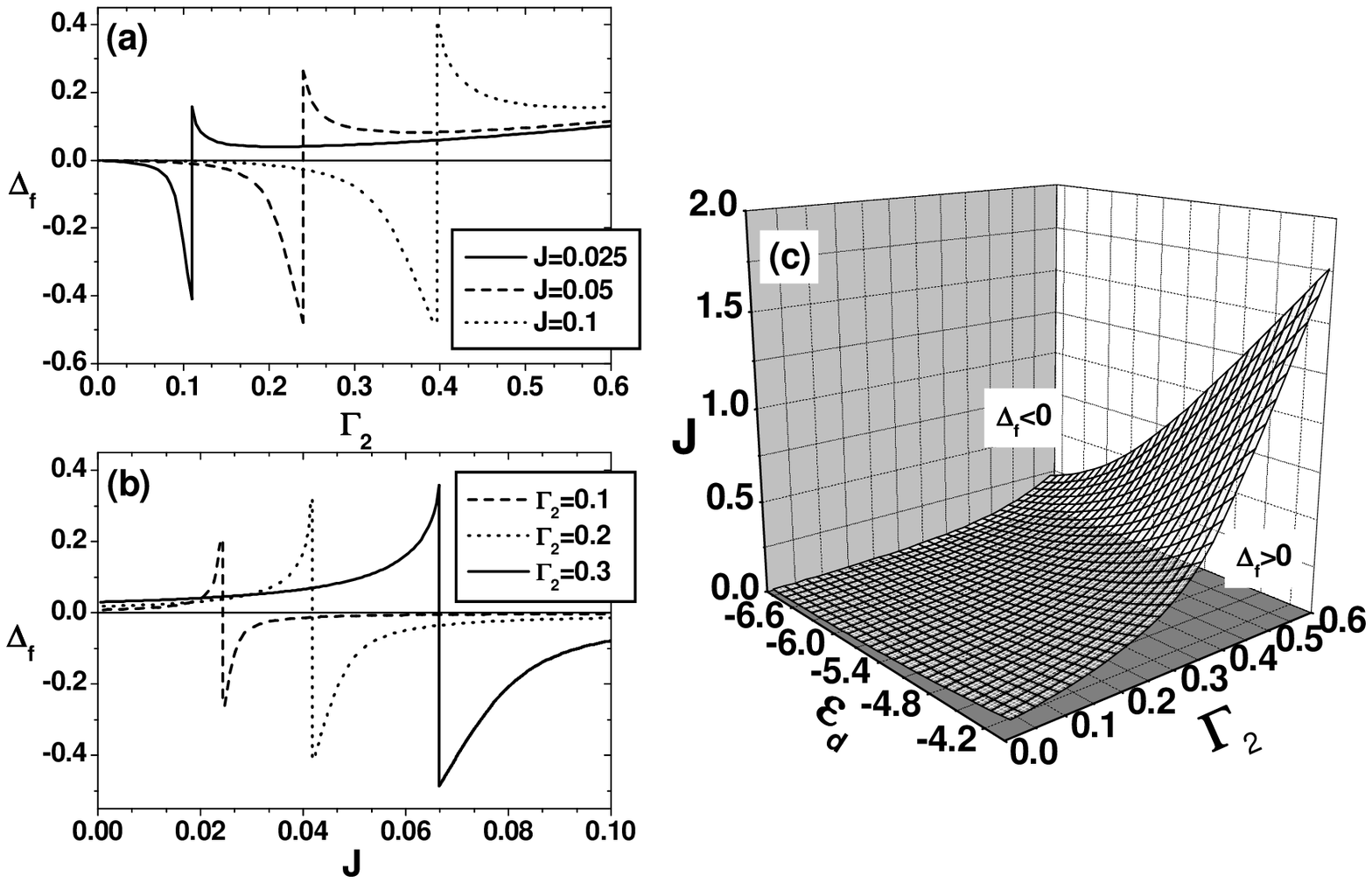}
\end{center}
\caption{The order parameter $\Delta _{f}$ in the Kondo regime ($\protect%
\epsilon _{d}=-6$) as functions of the asymmetry ratio of the dot-lead
couplings $\Gamma _{2}$ in (a) and the ISS $J$ in (b). The ground state
phase diagram is given in (c). $\Gamma _{1}=1$ is as the energy unit.}
\label{order_parameter}
\end{figure}

\begin{figure}[tbp]
\begin{center}
\includegraphics[width=3.2in]{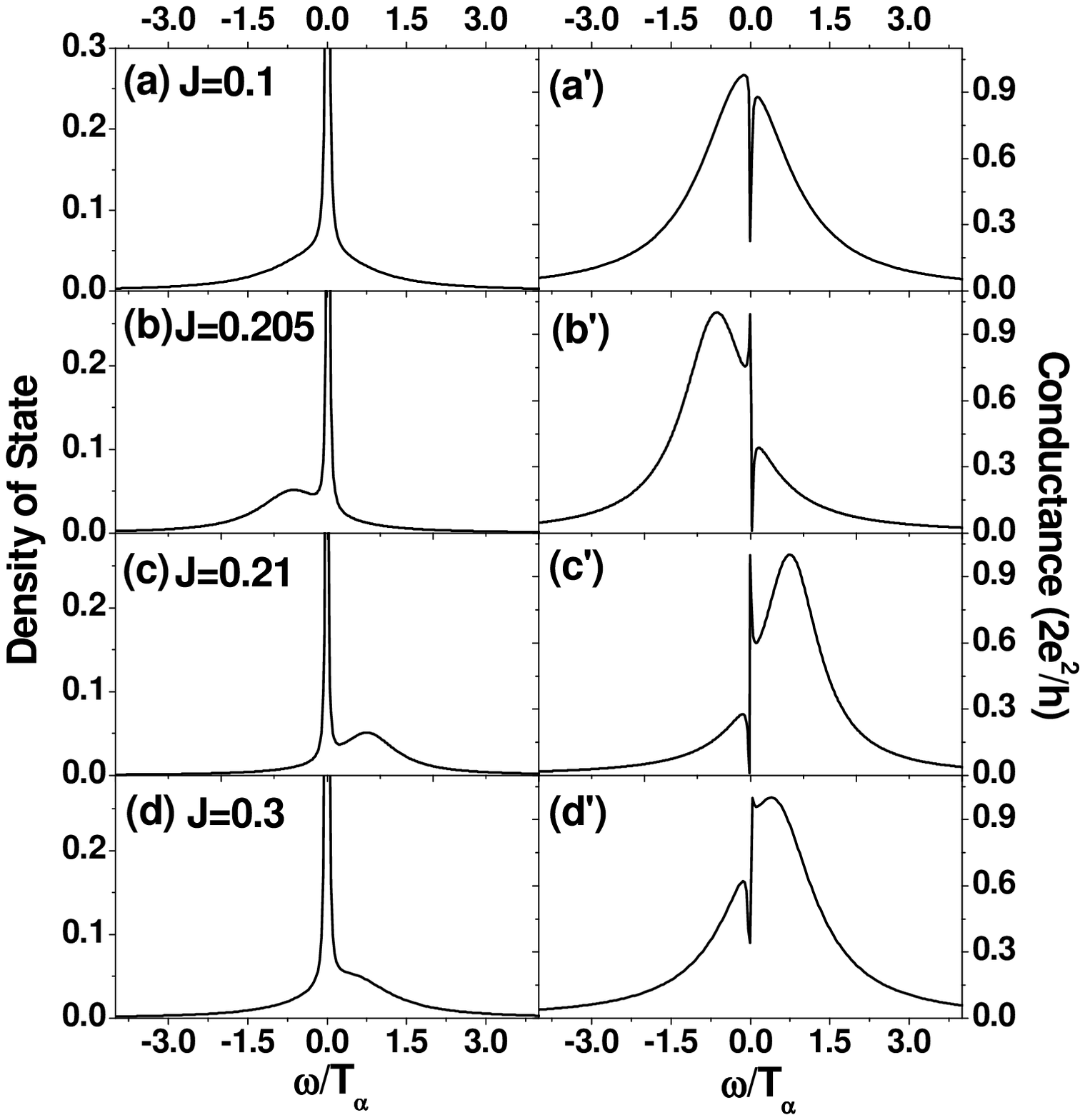}
\end{center}
\caption{The dot DOS and the corresponding linear conductance for different
interdot AF spin superexchange $J$ with a given asymmetry ratio of the
dot-lead couplings $\Gamma _{2}=0.6$ and degenerate dot energy levels $%
\protect\epsilon _{d}=-6$.}
\label{DOS}
\end{figure}

In the Coulomb blockade regime for each dot, the degenerate dot energy level
is set by $\epsilon _{d}=-6$, far below the Fermi energy. We find that $%
T_{\alpha }$ grows as increasing $\Gamma _{2}$, while $T_{\beta }$ is a
small constant ($T_{\beta }<T_{\alpha }$). For a given value of $\Gamma _{2}$%
, both $T_{\alpha }$ and $T_{\beta }$ are almost independent of $J$. In the
parallel-coupled DQD with a tunable ISS, the electron occupation on each dot 
\textit{strongly} depends on parameters $\Gamma _{2}$ and $J$. Given a
dot-lead coupling asymmetry ratio $\Gamma _{2}$, we find that the MF value
of $\Delta _{f}$ grows as $J$ increasing and changes sign at a critical
value $J_{c}$. When $\Delta _{f}$ reverses sign, the relative position of
the bonding and antibonding energies of the renormalized dot energy levels
are switched, \textit{i.e.}, the quantum spin states on QDs are swapped.
Similarly, for a fixed value of $J$, there also exists a critical value of
the dot-lead coupling asymmetry ratio ($0<\Gamma _{2,c}<1$), where $\Delta
_{f}$ changes sign. Such a sign change effect signals the existence of two
different regimes in the \textit{asymmetric} parallel-coupled DQD system. In
Fig.\ref{order_parameter}, we display EOP $\Delta _{f}$ as functions of the
ISS $J$ and the ratio of the dot-lead couplings $\Gamma _{2}$. Actually, the
dramatic effect of $\Delta _{f}$ is not sensitive to the parameter $\epsilon
_{d}$ as long as the condition $|\epsilon _{d}|>\Gamma _{2}$ is satisfied.
In Fig.\ref{order_parameter}c, a phase diagram is given in the
three-parameter space ($\epsilon _{d},\Gamma _{2},J$). 
Moreover, we evaluated the dot electron DOS for $\epsilon _{d}=-6$ and $%
\Gamma _{2}=0.6$ with different ISS, shown in the left column of Fig.\ref%
{DOS}. When $J=0.1$, $\Delta _{f}$ has a very small value, and the entangled
bonding and antibonding resonances are very close to each other and hardly
distinguishable in Fig.\ref{DOS}a. For $J=$ $0.205$, $\Delta _{f}$ reaches
its largest value and the entangled bonding resonance appears below the
Fermi energy (Fig.\ref{DOS}b). When $J=0.210$, $\Delta _{f}$ changes sign
and the bonding resonance state swaps its position from below to above the
Fermi level. At $J=0.3$, the bonding and antibonding resonances emerge
again. In this process, a sharp and narrow entangled resonance always stands
close to the Fermi level.

\begin{figure}[tbp]
\begin{center}
\includegraphics[width=3.2in]{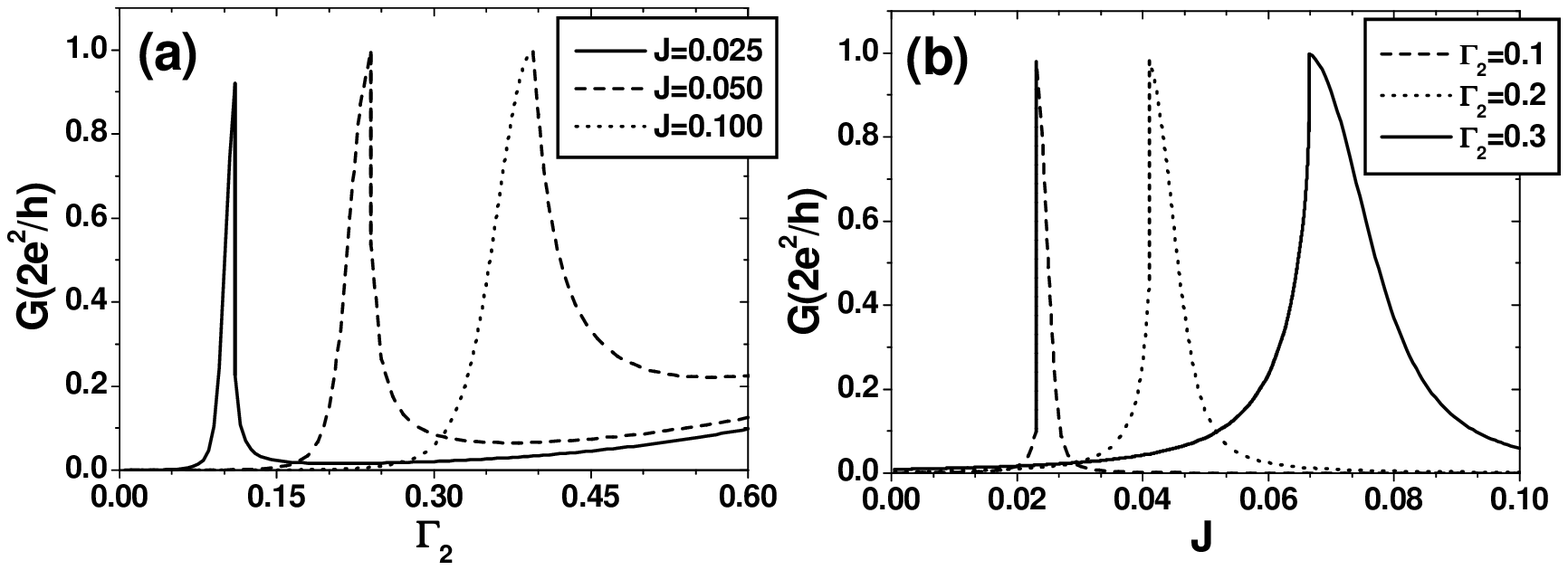}
\end{center}
\caption{The zero-bias differential conductance as functions of the
asymmetry ratio of coupling parameters $\Gamma_2$ and the ISS $J$ with
degenerate dot energy level $\protect\epsilon_d=-6$.}
\label{conductance}
\end{figure}

From the Landauer formula, the linear conductance can also be calculated 
\begin{equation}
G(\omega )=\frac{2e^{2}}{h}\mathrm{Tr}\left[ \mathbf{G}_{f}(\omega -i0^{+})%
\mathbf{\Gamma }_{R}\mathbf{G}_{f}(\omega +i0^{+})\mathbf{\Gamma }_{L}\right]
,
\end{equation}%
where $\mathbf{G}_{f}(\omega +i0^{+})$ and $\mathbf{G}_{f}(\omega -i0^{+})$
represent the retarded and advanced Green's functions, and $\mathbf{\Gamma }%
_{R}=2b_{0}^{2}\left( 
\begin{array}{cc}
\Gamma _{2}, & \sqrt{\Gamma _{1}\Gamma _{2}} \\ 
\sqrt{\Gamma _{1}\Gamma _{2}}, & \Gamma _{1}%
\end{array}%
\right) $ and $\mathbf{\Gamma }_{L}=2b_{0}^{2}\left( 
\begin{array}{cc}
\Gamma _{1}, & \sqrt{\Gamma _{1}\Gamma _{2}} \\ 
\sqrt{\Gamma _{1}\Gamma _{2}}, & \Gamma _{2}%
\end{array}%
\right) $ are two matrices related to the asymmetric dot-lead couplings. The
corresponding results are delineated in the right column of Fig.\ref{DOS}.
For the intermediate couplings of $J$, the linear conductance displays a
Lorentzian conductance peak centered at the bonding energy and a Fano
resonance at the antibonding energy (Fig.\ref{DOS}b'). The latter arises due
to the presence of bound states of DQD embedded in the conduction band
continuum. However, when the EOP $\Delta _{f}$ changes sign, these two
characteristic line shapes are switched (seen in Fig.\ref{DOS}c'). For a
smaller value of $J$ in Fig.\ref{DOS}a', the Fano resonance with a small
transmission is an anti-resonance at the Fermi energy, because of the
destructive quantum interference between different pathways through QDs. As $%
J$ is large in Fig.\ref{DOS}d', there is a progressive increase of the width
of the bonding resonance. For the serial-coupled ($\Gamma _{2}=0$) and
symmetric parallel-coupled ($\Gamma _{2}=\Gamma _{1}$) DQD, these swapping
features do not appear. 
To make contacts with experiments directly, we extract the zero-bias
differential conductance $G(\omega =0)$ as functions of the asymmetric
coupling parameters $\Gamma _{2}$ and ISS $J$ for a given value of $\epsilon
_{d}=-6$, displayed in Fig.\ref{conductance}a and \ref{conductance}b.
Surprisingly, we find that the differential conductances have maxima \textit{%
precisely} at the critical couplings for swap. The appearance of such sharp
conductance peaks can be explained by the majority occupation of the bonding
quasiparticle states across the Fermi level during the swap process. 
We should point out that the singlet-triplet transition discussed here is
very different from what was considered in multilevel dots \cite%
{hs-2002,hz-2003,pgh-2003,ka-2002}, where the coupling parameter symmetry
excludes completely the antibonding channel of the conduction electrons, so
the swap effect does not appear there. It is true that the quantum
fluctuations beyond the MF description may change detailed behavior near the
swap point, but the existence of the swap itself is a robust effect, because
two MF solutions with opposite sign of $\Delta _{f}$ are stable.

To conclude, in a degenerate parallel-coupled DQD with asymmetric parallel
couplings, two entangled bonding and antibonding resonances are formed close
to the Fermi energy in the Kondo regime of each dot. A swap effect between
two resonances has been found, leading to a sharp peak centered at the
critical coupling in the zero-bias differential conductance. To observe such
a peculiar effect in experiments, one has to fine tune the corresponding
gate voltage controlling the interdot electron hopping or the dot-lead
coupling asymmetry ratio in a parallel-coupled DQD system. Moreover, this
effect will lead to a practical and reliable mechanism to construct the
quantum gate for the future quantum computing.

G.M. Zhang is supported by NSF-China (Grant Nos. 10074036 and 10125418) and
the Special Fund for Major State Basic Research Projects of China (Grant No.
G2000067107), while L. Yu is supported by NSF-China grant No. 90203006.

\end{document}